# Detecting total hip replacement prosthesis design on preoperative radiographs using deep convolutional neural network

Alireza Borjali[a,b], Antonia F. Chen[c], Orhun K. Muratoglu[a,b], Mohammad A. Morid[d], *Kartik M. Varadarajan[a,b]

[a] Department of Orthopaedic, Harris Orthopaedic Laboratory, Massachusetts General Hospital, Boston, MA
[b] Department of Orthopaedic Surgery, Harvard Medical School, Boston, MA
[c] Department of Orthopaedic Surgery, Brigham and Women's Hospital, Harvard Medical School, Boston, MA
[d] Department of Information Systems and Analytics, Santa Clara University Leavey School of Business, Santa Clara, CA

*Corresponding author
Email: kmangudivaradarajan@mgh.harvard.edu
Phone: 617-643-3850
Address: 55 Fruit Street, GRJ-12-1223
Boston, MA, 02214

**RUNNING TITLE:** Detecting THR Implant With CNN

**AUTHOR CONTRIBUTIONS STATEMENT:** AB, AM, and KV designed the study. AB performed the study. AB, KV, AC, and OM performed the analysis and interpretation of the results. Each author has contributed to the writing and revising the manuscript. All authors have read and approved the final submitted manuscript.

**ABSTRACT**

Identifying the design of a failed implant is a key step in preoperative planning of revision total joint arthroplasty. Manual identification of the implant design from radiographic images is time consuming and prone to error. Failure to identify the implant design preoperatively can lead to increased operating room time, more complex surgery, increased blood loss, increased bone loss, increased recovery time, and overall increased healthcare costs. In this study, we present a novel, fully automatic and interpretable approach to identify the design of total hip replacement (THR) implants from plain radiographs using deep convolutional neural network (CNN). CNN achieved 100% accuracy in identification of three commonly used THR implant designs. Such CNN can be used to automatically identify the design of a failed THR implant preoperatively in just a few seconds, saving time and improving the identification accuracy. This can potentially improve patient outcomes, free practitioners time, and reduce healthcare costs.

**1. INTRODUCTION**

Identification of failed implant design is one of the key steps in preoperative planning of revision total joint arthroplasty. Wilson *et al.* conducted a detailed survey of the practices and challenges surrounding preoperative identification of failed hip and knee components. They determined that the use of patient x-rays, followed by hospital operative records, office records, operative dictation reports and implant sheet/labels, were the top five methods used by orthopedic surgeons for preoperative component identification [1]. The median time spent by surgeons and their staff on device identification was 20 min per case and 30 mins per case, respectively, for an estimated 41 hours per year per surgeon. The opportunity cost associated with this was estimated to reach $27.4 million by 2030 based on surgeon Medicare reimbursement for Evaluation and Management (E/M) 99213 [2]. Further, in an estimated 10% of the cases surgeons could not identify the device pre-operatively, and in about 2% of cases, they failed to identify the device intra-operatively. The failure to identify the implant preoperatively was perceived to result in $\geq 2$ implants brought to the case, increased operating room time, more complex surgery, increased blood loss, increased bone loss, increased recovery time, and increased healthcare costs [1,2]. This demonstrates both the inefficiencies associated with routine identification of failed components prior to revision arthroplasty, as well as the potentially significant impact on patient health and overall cost of care when such identification cannot be accomplished.

A key barrier to accurate device identification is the use of non-standardized and manual methods for documentation within medical records, resulting in incomplete device information [1]. This was indicated as a barrier in component identification at least some of the time by 88% of respondents, and always/most of the time by 27% of respondents, in the survey of orthopaedic surgeons conducted by Wilson *et al*. In recent years, there have been some important regulatory efforts to improve implant identification and traceability. In 2014, the Food and Drug Administration (FDA) issued a new rule requiring manufacturers to label medical devices with unique device identifiers (UDIs) and established the publicly accessible Global Unique Device Identifier Database (GUDID)[2]. Furthermore, the Office of the National Coordinator (ONC) for Health Information Technology and Centers for Medicare and Medicaid Services (CMS) mandated the inclusion of UDI in the patients' medical records beginning in 2015 and 2018, respectively [3]. These developments should dramatically improve the ability to accurately identify failed implant components in the future. However, the large numbers of patients already implanted with total joint replacements in the preceding decades would not benefit from these developments. As of 2014, over 10 million patients in the United States (US) are living with a hip or knee replacement [4]. A large proportion of patients undergoing revision do so at an institution different from where they had their primary procedure (over 40% by 3 years post primary procedure), further complicating the device identification process [5]. Thus, the challenge of accurate device identification remains for many patients in the US and around the world.

Plain radiography is the cornerstone of pre-operative planning for revision procedures, especially for identifying failed components. In the past, practitioners have strived to aid manual identification of device design through establishing reference radiographic image galleries [6–8]. We hypothesized that modern deep learning based artificial intelligence algorithms could be trained to automatically identify hip implant design from radiographic images. Deep learning (DL) is a sub-set of the broader family of artificial intelligence (AI) or machine learning methods that leverages artificial neural networks for automatic object-detection and classification of imaging datasets. DL methods have already been applied to plain film radiographs with high degrees of success for identification of wrist, elbow, humerus, ankle and hip fractures, classification of proximal humerus and hip fracture types, detecting presence and type of arthroplasty, staging knee osteoarthritis (OA) severity, etc [9–17]. Performance of the machines in these cases was typically on-par with trained surgeons and radiologists, and superior to general practitioners. Thus, for the first time in human history, machines are able to replicate and, in many instances, surpass the visual pattern recognition capabilities of humans. Recently, we successfully trained a DL algorithm to assess aseptic loosening status of total hip replacement (THR) implants via radiographs [18]. The ability to identify device design on radiographs could be an added capability provided by such algorithms, further adding to the armament of tools available to the clinical practitioner. With this background in mind, the purpose of this study was to determine: 1) whether a deep convolutional neural network (CNN) could be trained to provide automated identification of THR implant designs from radiographic images, and 2) whether the decision-making process of the CNN could be visualized to build confidence in machine prediction.

## 2. METHODS

This study provides level III evidence. After acquiring institutional review board (IRB) approval, we conducted a retrospective study using previously collected imaging data at a single institution during 1/2018-12/2018. We evaluated post-surgery anteroposterior (AP) hip x-rays of 252 primary THR patients with three different commonly used implant designs. These three designs accounted for the majority of the implant designs (53%) that were used in the institution during this study period. We used the design of the stem recorded in the primary surgery operative note as the ground truth label of the x-ray images. The stem designs used were: 1) Accolade II, manufactured by Stryker corporation (Mahwah, NJ, USA). 2) Corail, manufactured by DePuy Synthes companies (Warsaw, IN, USA), and 3) S-ROM, also manufactured by DePuy Synthes companies (Warsaw, IN, USA). All Corail stems included in this study were collared. Table 1 shows the THR patient information and the distribution of implant designs.

**Table 1** Total hip replacement (THR) patient information

| THR implant design | Accolade II | Corail | S-ROM |
|---|---|---|---|
| Number of patients | 130 | 93 | 29 |
| Age (mean ± standard deviation) | 62.3 ± 13.0 | 66.8 ± 13.0 | 57.2 ± 16.3 |
| Male | 67 (51.5%) | 41 (44.1%) | 18 (62.1%) |
| Female | 63 (48.5%) | 52 (55.9%) | 11 (37.9%) |

Fig. 1 shows an x-ray example of these three different THR implant designs. While the acetabular component of these implant system had generally similar designs, the stem designs varied more substantially. The medial offset, vertical height, neck-shaft angle, stem length and distal relief of the stems were different. Furthermore, the Corail stems included in this study had collars and the S-ROM stems had modular metaphyseal sleeves.

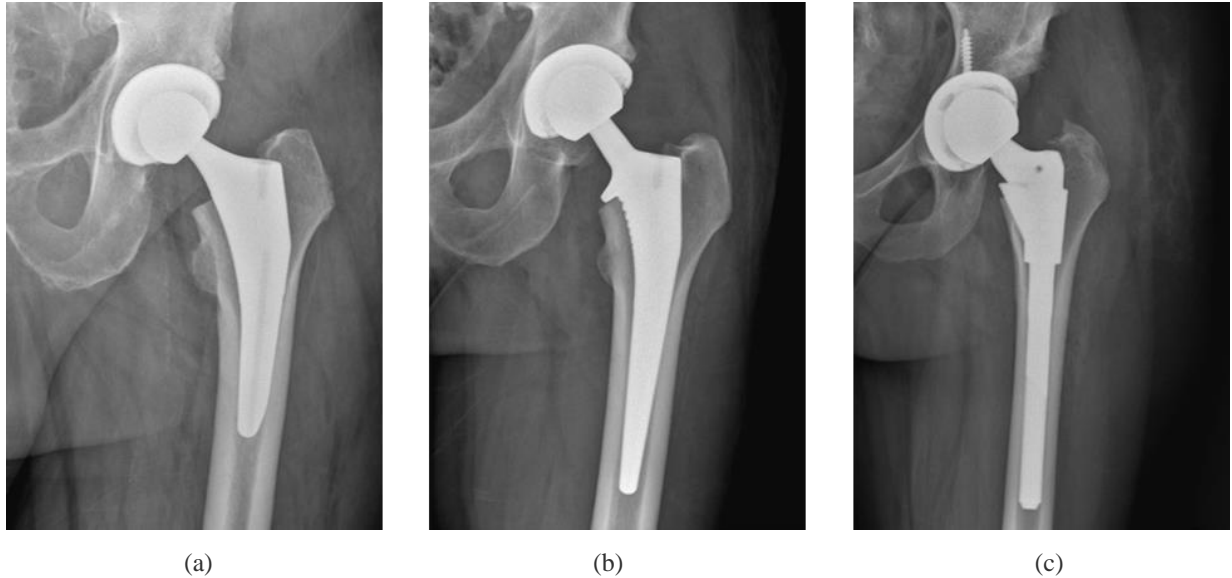

(a)            (b)            (c)

**Fig 1.** Three designs of commonly used THR implants: (a) Accolade II, (b) Corail, and (c) S-ROM

We removed the annotations and anonymized the x-rays by separating the embedded patient's information in the Digital Imaging and Communications in Medicine (DICOM) files from the x-ray images prior to any analysis. We randomly split the x-rays dataset into train, validation, and final test subsets with 80:10:10 split ratio[19]. We maintained the split ratio among all the implant designs x-rays to make sure that each design had representative x-rays in the training, validation, and test subsets. We used the training subset with online data augmentation to train the CNN. Online data augmentation created new data during the training process by applying minor changes on the base training subset. These minor changes to the base train subset helped to account for real-world variation of x-rays such as slight variation in orientation, magnification, and hip positioning. Data augmentation increased the invariance of the CNN to the real-world variation and reduced the chance of overfitting on the train subset. Hyper-parameters including learning rate, learning rate decay, regularization, batch size, and number of epochs were optimized iteratively on the validation dataset using a grid search strategy. We adopted a CNN that was developed for non-medical image classification and modified for our application. This method is referred to as "transfer learning", where a CNN initially developed for one specific application is "transferred" to be used in another application. We used DenseNet [20] CNN architecture and replaced the final classifier nodes with three nodes for categorizing the x-rays into the three THR implant designs. In a DenseNet CNN architecture, each layer is connected to every other layer through feed-forward connections as opposed to a traditional CNN, where there is only one connection between sequential layers. This architecture achieves high performance utilizing less computational power compared to the other state-of-the-art CNN architectures [20].

We implemented two different weight initialization methods and compared the results: (1) in the first method, we initialized the weight with a random Gaussian distribution and re-trained the CNN on x-ray images (referred to as "re-trained CNN" henceforth), (2) in the second method, we used the weights from a pre-trained CNN on the large ImageNet [21] database and trained the classifier on x-ray images (referred to as "pre-trained" CNN henceforth). ImageNet is a database of non-medical images consisting of more than 14 million entries belonging to 27 high-level categories such as "bird", "flower", and "food", to name a few.

We implemented image-specific saliency maps to shed light on the CNN decision-making process. Image-specific saliency map ranked all the pixels of a given x-ray based on their influence on the CNN's classification output. Saliency maps helped to visualize where the network was "looking" at to make a classification, thus making the results more interpretable [22]. Making the CNN more interpretable can increase the confidence in its output and might result in new insights by revealing any hidden knowledge that the CNN potentially leverages to make its decisions.

The CNN was trained using Adam optimizer (initial learning rate = 0.001, beta 1 = 0.9, beta 2 = 0.999, epsilon = 1 e-8), with a batch size of 5 for 350 epochs. The CNN was implemented using Tensorflow (Keras) on a workstation comprised of an Intel(R) Xeon(R) Gold 6128 processor, 64GB of DDR4 RAM and a NVIDIA Quadro P5000 graphic card. After achieving optimum performance on the validation subset, the CNN was tested on the holdout test subset and the results were reported as the CNN output performance.

## 3. RESULTS

Fig. 2 shows the training loss and validation accuracy as a function of the number of epochs for both re-trained (Fig. 2 [a]) and pre-trained (Fig. 2 [b]) CNNs classifying the x-rays into three designs of THR implant.

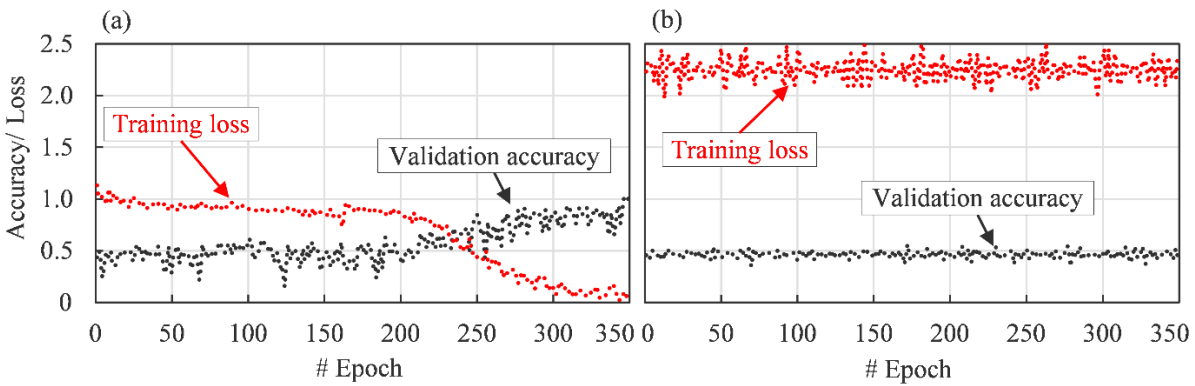

**Fig. 2** Training loss and validation accuracy as a function of number of epochs for (a) pre-trained convolutional neural network (CNN) and (b) re-trained CNN

Fig. 2 (a) shows that the re-trained CNN achieved 100% accuracy on the validation subset after 350 epochs. On the other hand, Fig. 2 (b) shows that the training loss for the pre-trained CNN did not decrease as a function of training epochs, and the validation accuracy remains almost constant. This shows that the pre-trained CNN never learned how to categorize the x-rays.

Fig. 3 shows an example x-ray of each THR implant designs and the corresponding re-trained CNN's saliency map with insets magnifying the tip of the stem for all three implants, the collar in Corail, and the metaphyseal sleeve of the S-ROM . Colored regions in the saliency map indicate most influential regions on the re-trained CNN's performance, where red denotes higher relative influence than blue. Re-trained CNN classified all three THR implant designs with 100% accuracy on the test subset (25 x-rays), which had never been included in the training or the validation process. It took the re-trained CNN about 2 minutes to classify all the x-rays in the test subsets, spending roughly 5 seconds per x-ray. The saliency maps showed that the re-trained CNN detected the design of implant by "looking" at the acetabular cup used with Accolade II and S-ROM stems, the collar in Corail, the metaphyseal sleeve of S-ROM, and the tip of the stem for all three implants. The fact that the re-trained CNN considered the tip of the stem in its decision making, was surprising at first glance, but a closer look showed that the tip of the stem's geometry could be used as a unique feature to identify the design of the implant. The differences in tip of the stem's geometries were significant enough that at the second glance we could differentiate between the implant designs by just looking at tip of the stems. This is an example of machine giving the human observer a new insight about a problem and pointing out new important features in a radiograph that humans might miss or not consider in their assessment. It is important to mention here that the machine was never directly programmed to look at these regions, and it "learned" to look at these regions and ignored the rest of the image and background noise. On the other hand, pre-trained CNN did not learn how to classify the x-rays.

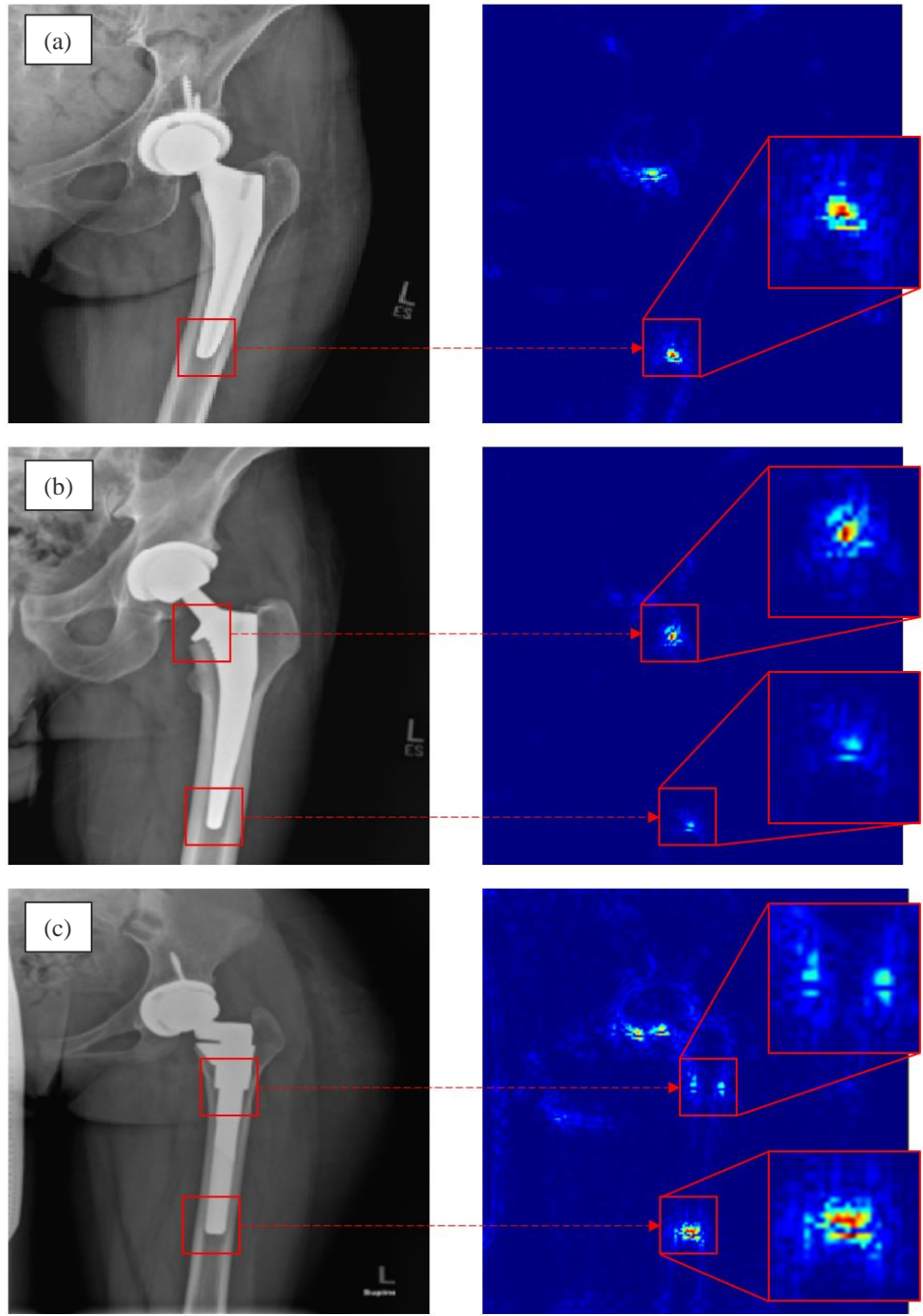

**Fig. 3** Three total hip replacement (THR) implant designs and the corresponding re-trained convolutional neural network (CNN) saliency maps showing (a) Accolade II with an inset magnifying the tip of the stem, (b) Corail with insets magnifying the tip of the stem and the collar, and (c) S-ROM with insets magnifying the tip of the stem and the metaphyseal sleeve. Saliency maps indicate most influential regions on the re-trained CNN's performance where red denotes higher relative influence than blue.

Fig. 4 shows an example Accolade II x-ray (Fig. 4[a]) and the corresponding saliency maps for re-trained (Fig. 4 [b]) and pre-trained (Fig. 4 [c]) CNNs. Fig. 4 shows that while the re-trained CNN focused at the implant stem tip and the acetabular cup and ignored the background noise, the pre-trained CNN struggled to find relative features and was influenced by the background and other irrelevant features trying and failing to detect the implant design.

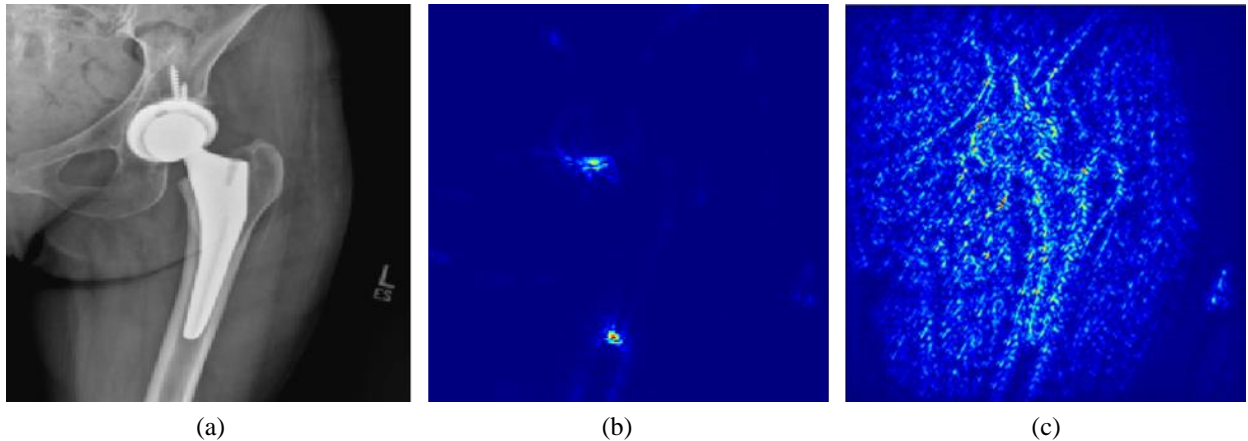

(a)　　　　　　　　　　　(b)　　　　　　　　　　　(c)

**Fig. 4** An example Accolade II THR implant showing (a) x-ray, (b) the corresponding saliency map for re-trained CNN, and (c) pre-trained CNN

## 4. DISCUSSION

In this study, we implemented DL method to automatically detect the design of THR implants from plain film radiographs. Other studies have applied DL methods on plain film radiographs for various orthopedic applications [9–17]. However, to the best of our knowledge, this is the first study to apply DL method to automatically detect THR implant designs. We used transfer learning method to implement pre-trained and re-trained CNNs on AP hip x-rays of 252 patients to teach the algorithms to detect three different designs of commonly used THR implants. We evaluated both CNNs on the test subset, which was isolated from the training and validation process. The re-trained CNN achieved 100% accuracy classifying all the x-rays in the test subset correctly, thus proving our hypothesis that a CNN can be trained to provide automated identification of THR implant designs from radiographic images. This shows the potential of integrating CNN in orthopaedic care where it can be used to automatically identify the design of THR implants in a few seconds, saving time and improving the detection accuracy of practitioners. This can potentially have a significant impact on improving patient's health and decreasing overall cost of revision THR procedures by identifying the right components to replace.

We also showed potential challenges with implementation of transfer learning for categorical classification specific to orthopaedics. We showed that a CNN pre-trained on non-medical images may not always be useful for orthopaedic categorical classification, as the pre-trained CNN failed to classify the x-rays in the current application. This shows that pre-training a CNN on non-medical images, e.g. to classify those images into categories such as "bird", "flower", and "food", may not be directly transferrable to an orthopedic application. This is likely because the image features that differentiate Accolade II, Corail, and S-ROM THR implants are significantly different from the image features that differentiate between non-medical image categories such as "bird", "flower", and "food". Previously, pre-trained CNN has been mainly used in orthopaedics for binary image classification as opposed to categorical classification [12,13,15,] and generally no direct evidence has been provided to support the specific choice as being the optimal approach We show that the choice of pre-trained or re-trained CNN might be application dependent; however, more research is required to further assess this.

We also used saliency maps to shed light on the decision-making process of the CNN. We showed that the re-trained CNN looked at new features to make a classification that the human observer might not notice at first glance. This was an example of AI pointing out new features, and hidden knowledge in the database that human might miss. While a few other studies [10,11,17] have used saliency maps to visualize the CNN's final output, the use of saliency maps as a tool for pointing out new features and hidden knowledge in the database, as presented here, is novel.

The primary limitation of this study is the size of the dataset. With a larger dataset in future studies, we can consider more THR implant designs including non-collared Corail stems (all Corail stems were collared in this study), and other designs with more subtle geometric differences. Furthermore, we only used one AP x-ray per patient, as opposed to clinical practice, where the clinician would have the benefit of additional views to identify the design of a THR implant.

In this study, we presented a novel, fully automatic and interpretable approach to identify the design of THR implants from AP hip x-ray using deep CNN. We intend to further develop this AI method to be able to identify more THR implants designs as well as other implants (e.g. total knee replacement implants) incorporating additional radiographic views. Automated and accurate identification of failed implant components may prove to be valuable addition to the tool-kit of revision arthroplasty surgeons.

## AKNOWLEDGEMENT

This study was supported by laboratory sundry funds, and no external funding was received. The authors have no financial disclosures relevant to this study.